\documentclass[12pt]{article}
\usepackage[margin=1in]{geometry}
\usepackage{amsmath}
\usepackage{graphicx}
\usepackage{float}
\usepackage{cite}
\usepackage{inputenc}
\usepackage{amssymb}
\usepackage{authblk}
\usepackage{xcolor}

\title{Non-Custodial Warped Extra Dimensions at the LHC?}
\author[1]{Barry M. Dillon}
\author[1]{Stephan J. Huber}
\affil[1]{\textit{\small{Department of Physics and Astronomy, University of Sussex, BN1 9QH Brighton, UK}}}
\date{\today}

\begin{document}
\maketitle
\begin{abstract}
With the prospect of improved Higgs measurements at the LHC and at proposed future colliders such as ILC, CLIC and TLEP we study the non-custodial Randall-Sundrum model with bulk SM fields and compare brane and bulk Higgs scenarios.  The latter bear resemblance to the well studied type III two-Higgs-doublet models.  
We compute the electroweak precision observables and argue that incalculable contributions to these, in the form of higher dimensional operators, could have an impact on the $T$-parameter.  This could potentially reduce the bound on the lowest Kaluza-Klein gauge boson masses to the $5$ TeV range, making them detectable at the LHC.  In a second part, we compute the misalignment between fermion masses and Yukawa couplings caused by vector-like Kaluza-Klein fermions in this setup.  The misalignment of the top Yukawa can easily reach $10\%$, making it observable at the high-luminosity LHC. Corrections to the bottom and tau Yukawa couplings can be at the percent level and detectable at ILC, CLIC or TLEP.
\end{abstract}

\newpage

\section{Introduction}
Due to their attractive model building features and rich phenomenology, warped extra dimensional models have been studied extensively for fifteen years.  The first proposal of such a model was by Randall and Sundrum (RS) in 1999 \cite{Randall:1999ee}, and consisted of an AdS space mapped onto an $S_1/Z_2$ orbifold bounded by two 3-branes.  The AdS geometry imposes an exponential hierarchy in energy scales between the two branes, thus, with all standard model (SM) fields residing on the low energy (IR) brane and with a suitable choice of parameters, this model offers a simple and natural solution to the hierarchy problem.  Studying perturbations to this metric reveals that the graviton zero mode is localised towards the high energy (UV) brane and hence the interaction of gravity with SM fields is naturally weak.  In addition to this, it was shown that the size of the extra dimension can be stabilised without fine-tuning using a bulk scalar field \cite{Goldberger:1999uk}.  

Extending this model to have the SM fields propagating in the bulk provides a more interesting phenomenology, but also more stringent constraints on model parameters.  The most striking feature of these models is the presence of Kaluza-Klein (KK) modes in the 4D effective theory, of which the zero modes are identified with the SM particles.  These arise  due to the compactification of the bulk fields.  The masses of scalar, gauge and fermion KK modes represent a scale of new physics in the effective model which is expected to be in the TeV range.  

In addition to solving the hierarchy problem, these models are motivated by explaining the fermion mass hierarchy \cite{ArkaniHamed:1999dc,Huber:2000ie}, new mechanisms for supersymmetry breaking \cite{Gherghetta:2000qt,Luty:1999cz,Burgess:2006mn}, and by composite Higgs models where the AdS background is dual to a strongly coupled 4D theory through the AdS/CFT correspondence \cite{Agashe:2004rs,Contino:2006nn} (see ref.~\cite{Contino:2010rs} for a recent review). 

In this paper revisit the case of a bulk Higgs field. We first look at how the presence of the Higgs KK modes induce electroweak corrections to the SM.  We then propose that higher dimensional operators in the 5D theory may reduce these constraints on the new physics scale.  Lastly we study the effects of a bulk Higgs on the Yukawa couplings. We find a large correction to the SM Yukawa couplings for heavy fermions which has an interesting dependence on the Higgs localisation. Although these deviations may be detectable at future collider experiments, such as the high-luminosity LHC, at present they do not place additional constraints on the bulk Higgs scenario. 

In section 2 we give an overview of the treatment of bulk scalar, gauge and fermion fields in the RS model.  We aim to present general results which are used throughout the paper, but we also include a discussion on the fine-tuning of scalar fields in the model.  For a more detailed analysis on bulk fields see \cite{Gherghetta:2000qt,Davoudiasl:1999tf,Huber:2000fh}, and specifically for a bulk Higgs see \cite{Davoudiasl:2005uu,Cacciapaglia:2006mz}.  It was first thought that models with a  bulk Higgs field required a large fine-tuning to obtain an electroweak (EW) scale zero mode with TeV scale KK modes \cite{Chang:1999nh}.  However, it was realised that if the bulk Higgs is localised towards the IR then one can naturally accommodate a light Higgs in the spectrum. 

In section 3 we look at the Higgs potential in 5D and study the effects of bulk and brane quartic terms.  We find that with a bulk quartic term the KK Higgs modes are more decoupled from the zero mode than with a brane quartic term.  The higher modes in the Higgs potential acquire v.e.v.'s and give additional mass to the SM fields, we find the effect this has on the Higgs couplings and particle masses to be too small for detection until we have a sub percent experimental accuracy on the Higgs couplings to gauge bosons and fermions.  An interesting observation which we discuss is that these effective theories may be viewed as multiple Higgs doublet models.

Constraints on the EW sector of these models are studied via the Peskin-Takeuchi parameters $S$, $T$ and $U$ \cite{Peskin:1991sw}.  In section 4 we calculate these parameters for our model.  The largest experimental bound comes from the $T$ parameter.  We confirm that with a brane Higgs the lower bound on KK gauge boson masses is about 15 TeV, and with a bulk Higgs this is reduced to around 8 TeV.  One way of reducing these stringent constraints is to extend the bulk gauge symmetry such that the KK gauge bosons in the effective theory preserve the custodial symmetry after electroweak breaking \cite{Agashe:2003zs,Carena:2007ua,Casagrande:2010si}.  Another way is to introduce a scalar field which back-reacts on the metric causing a departure from AdS in the IR \cite{Falkowski:2008fz,Cabrer:2010si,Cabrer:2011fb,Cabrer:2011mw,Carmona:2011ib}.  These mechanisms typically result in a lower bound of about 3 TeV.  Similar results can be obtained by introducing large brane kinetic terms for the gauge bosons \cite{Carena:2002dz} or by extending the space-time to include more than 5 dimensions \cite{Archer:2010bm,Archer:2010hh}.  Having more than 5 dimensions may allow for a reduction in constraints via volume suppression in the IR of the extra dimension.  We do not consider these extensions.  In the SM there is a set of dimension-6 operators contributing to the EW parameters. We promote these to 5D operators and study their effects.  The only one with a sizeable contribution is the 5D dimension-8 operator contributing to the $T$ parameter.  Assuming a mild cancellation, we find that this effect could possibly provide considerable reductions in the $M_{KK}$ bound, allowing KK resonances around 5 TeV, i.e.~within the range of LHC.

An exciting aspect of future collider experiments is the increased precision on top quark measurements.  Being the heaviest particle in the standard model, corrections to its properties from KK modes will generally be large.  The top quark mass is already well measured with the error being sub-percent.  However, measurements of the top Yukawa coupling still leave a lot of room for new physics, and the proposed future colliders could dramatically close this gap.  The precision forecasts from ILC \cite{Peskin:2012we,Asner:2013psa,Baer:2013cma}, CLIC \cite{Linssen:2012hp,Accomando:2004sz} and TLEP \cite{Gomez-Ceballos:2013zzn} state that they could achieve a precision $<5\%$ on the bottom and tau Yukawa couplings, and precision forecasts for the high luminosity LHC \cite{CMS:2013xfa,ATLAS:2013hta} indicate that they could achieve the same precision for the top quark.  In light of this, section 5 focuses on the misalignment of the fermion Yukawa couplings due to mixing with KK fermions. The largest of these effects is by far with the top quark, for which we find deviations from the SM could be as large as $\sim10\%$ for a bulk Higgs.  Similar calculations were done in \cite{delAguila:2000fg} for a brane Higgs.  We find some differences between the bulk and brane Higgs cases here.  One important difference is the reduced bound on the KK fermion scale, and another is the introduction of a new coupling not present in Brane Higgs scenarios.  Together, we find that these result in a larger Yukawa corrections for a bulk Higgs. While these are sizeable deviations from the SM, they currently do not lead to additional bounds beyond that from electroweak observables. KK resonances may therefore be indeed observable at LHC in the bulk Higgs setup.

\section{Bulk fields in Randall-Sundrum}
The Randall-Sundrum background is defined by the non-factorizable metric \cite{Randall:1999ee}:
\begin{equation}
ds^2=e^{-2k|y|}\eta_{\mu\nu}dx^{\mu}dx^{\nu}-dy^2 .
\end{equation}
The 4D metric is $\eta_{\mu \nu}=$diag$(1,-1,-1,-1)$, $k$ is the AdS curvature, and $y$ defines the position along the extra dimension.  The extra dimension is bounded by two 3-branes in the UV ($y=0$) and in the IR ($y=L$).  The length of the extra dimension, $L$, is assumed to be $\mathcal{O}(11 \hspace{1mm} \pi k^{-1})$ and is the free parameter which determines the new physics scale.  The AdS curvature is related to the fundamental 5D mass scale $M_5$ by
\begin{equation}
\kappa = \frac{k}{M_5}.
\end{equation}
This is discussed in more detail in \cite{Davoudiasl:1999jd} where the authors use $0.01\leq\kappa\leq1$ for their phenomenological analysis.  However, at the larger end of this range higher derivative corrections to the gravitational action will become important, rendering the derivation of the metric (1) unreliable.

\subsection{Scalar fields}
We write the action for the 5D scalar field as

\begin{equation}
S_{\Phi}=\int d^4x \int_0^Ldy\hspace{1mm}\frac{1}{2}\sqrt{|g|}\left((\partial_M\Phi)^2-m_{\Phi}^2\Phi^2\right),
\end{equation}
where $M=\mu,y$ and $\sqrt{|g|}=e^{-4ky}$.  The 5D mass term consists of both bulk and brane terms such that
\begin{equation} \label{scalarmass}
m_{\Phi}^2=(b^2+\delta b^2)k^2-\delta(y)a^2k+\delta(y-L)(a^2+\delta a^2)k.
\end{equation}
We perform a Kaluza-Klein expansion
\begin{equation}\label{KKscalar}
\Phi(x,y)=\frac{1}{\sqrt{L}}\sum_n \Phi_{n}(x)f_n(y)
\end{equation}
on the scalar field.

In the 4D spectrum, a massless zero mode exists if $\delta b^2$ and $\delta a^2$ vanish and the bulk and brane mass terms must are related by \cite{Gherghetta:2000kr,Huber:2002np}
\begin{equation} \label{masslessscalarrelation}
b^2=a^2(a^2+4).
\end{equation}
In figure 1 we see that the minimum value of the bulk mass which permits a massless solution is $-4$, this is known as the Breitenlohner-Freedman bound \cite{Breitenlohner:1982bm}, $b^2<-4$ would result in an unstable AdS space.  
\begin{figure}[ht] 
\centerline{\includegraphics[scale=0.25]{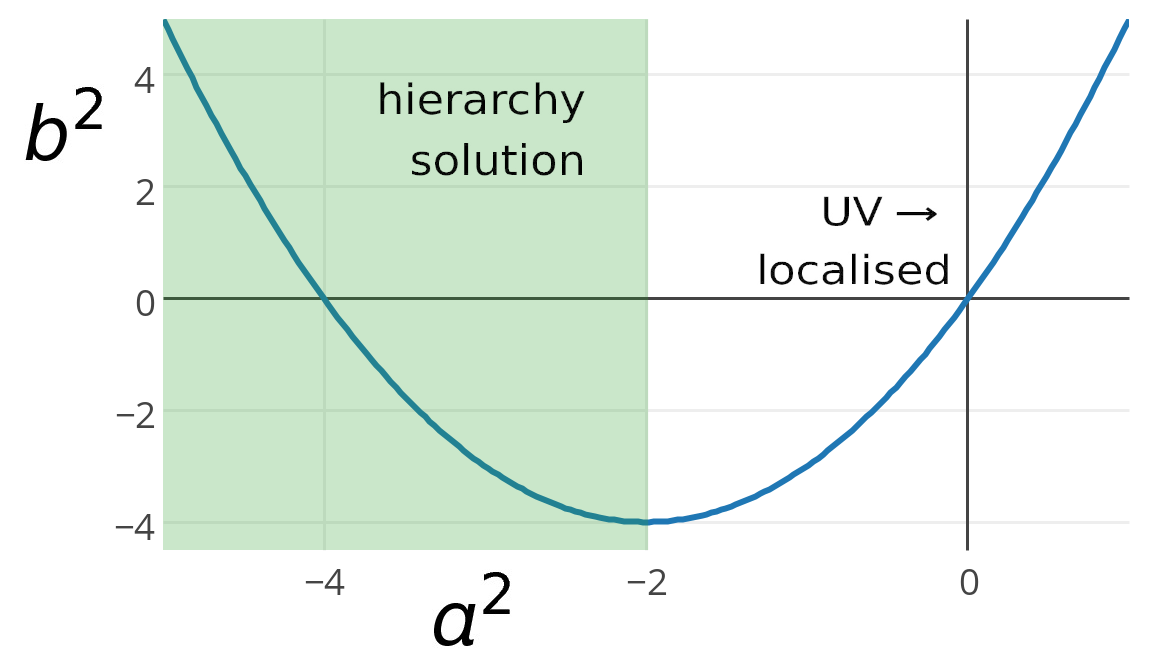}}
\caption{The solid line shows the relationship between the bulk and brane mass terms required to have a massless scalar mode of eq.~(\ref{masslessscalarrelation}).  The shaded region shows the parameter space for which the Higgs profile is sufficiently IR localised such that the hierarchy problem is resolved.}
\label{masslessscalar}
\end{figure}
By normalising the kinetic term and imposing the boundary conditions,
\begin{equation}\label{scalarbcs}
\begin{gathered}
f^{\prime}_n(L)=-a^2kf_n(L) \\
f^{\prime}_n(0)=-a^2kf_n(0), \\
\end{gathered}
\end{equation}
we find the zero mode profile to be,
\begin{equation} \label{scalarzero}
f_0(y)=\sqrt{\frac{2(1+a^2)kL}{1-e^{-2(1+a^2)kL}}}e^{-a^2ky}.
\end{equation}
The  parameter $a^2$ defines the localisation of the field in 5D and $a^2<0$ implies IR localisation. Along with this zero mode one obtains a tower of KK scalar fields with the following 5D profiles
\begin{equation}
f_n=\frac{e^{2ky}}{N_n}\left[J_{\alpha}\left(\frac{m_n}{k}e^{ky}\right)+\beta(m_n)Y_{\alpha}\left(\frac{m_n}{k}e^{ky}\right)\right],
\end{equation}
where $m_n$ is the mass of the $n$th mode, $\alpha=\sqrt{4+b^2}$, and $N_n$ is determined by the normalisation condition \cite{Gherghetta:2000qt}.  The constants $\beta(m_n)$ and the KK masses $m_n$ are determined by the boundary conditions and in the limit $kL>>1$ the KK masses can be approximated as
\begin{equation}
m_n\simeq \left(n+\frac{\alpha}{2}-\frac{3}{4}\right)\pi k e^{-kL} .
\end{equation}
Note that to obtain TeV-scale resonances we require that $kL\sim35$.

Switching on the mass perturbations $\delta b^2$ and $\delta a^2$ introduces a mixing between the KK modes of eq.~(\ref{KKscalar}).  We could also have a mass perturbation on the UV brane but that parameter is redundant.    
The effective action for our scalar can be written as
\begin{equation} \label{scalaraction}
S=\int d^4x \sum_{mn}\frac{1}{2} \left( (\partial_{\mu}\Phi_n)^2-m_n^2\Phi_n^2-\delta m_{mn}^2\Phi_m\Phi_n\right),
\end{equation}
where $\Phi_n$ are the 4D fields with wave functions $f_n(y)$ respecting the boundary conditions of eq.~(\ref{scalarbcs}).  The resulting contributions to the mass matrix are given by
\begin{equation} \label{scalarmasscorr}
\delta m_{mn}^2=\frac{\delta b^2k^2}{L}\int_0^Ldy\hspace{1mm}e^{-4ky}f_mf_n +\frac{\delta a^2k}{L}e^{-4kL}f_m(L)f_n(L).
\end{equation}
Once we turn on the mass perturbations $\delta b^2$ and $\delta a^2$ we turn on the mass mixings in the 4D effective theory. This requires us to diagonalize the mass matrix and in turn the zero mode becomes massive.  The effect on the masses of the higher modes is negligible.  With a slight tuning we can obtain a zero mode much lighter than the KK scale if the Higgs field is localised in the IR \cite{Davoudiasl:2005uu,Cacciapaglia:2006mz}.  Going to the mass eigenbasis we find that the zero mode mass is
\begin{equation}
m_0^2\simeq\delta m_{00}^2-\sum_{n=1}^{\infty}\frac{(\delta m_{0n}^2)^2}{m_n^2}.
\end{equation}
To adequately suppress the mass perturbations from $\delta b^2$ and $\delta a^2$ we need $a^2\leq-2$, see figure \ref{masslessscalar}.  The mass scale for the zero mode is set by $\delta m_{00}^2$. Setting $a^2=-2-x$ we find that
\begin{equation}
\delta m_{00}^2=\frac{2(1+x)k^2}{e^{2(1+x)kL}-1}\left[\frac{\delta b^2}{2x}\left(e^{2xkL}-1\right)+\delta a^2e^{2xkL}\right].
\end{equation}
Taking the limit $a^2\rightarrow-2$, this is found to be 
\begin{equation}
\delta m_{00}^2\simeq2(\delta b^2kL+\delta a^2)k^2e^{-2kL},
\end{equation}
and for $e^{2xkL}>>1$, i.e.~$x\gtrsim1/(kL)$
\begin{equation}
\delta m_{00}^2\simeq2(1+x)\left(\frac{\delta b^2}{2x}+\delta a^2\right)k^2e^{-2kL}.
\end{equation}
We see that, for $a^2=-2$, the bulk mass correction needs to be more tuned due to the $kL$ factor. However, as we move further towards the IR brane this enhancement of the bulk term quickly diminishes.  In all, we find that in order to have a zero mode at the electroweak scale we only require a percent level fine-tuning.  If $a^2>-2$, the mass corrections do not get the required suppression.  \newline

\subsection{Gauge fields}
The treatment of  gauge fields is similar to that of the scalar field.  We write the 5D action as
\begin{equation}
S_A=\int d^4x \int_0^Ldy\hspace{1mm}\sqrt{|g|}\left(-\frac{1}{4}F_{MN}F^{MN}\right).
\end{equation}
Working in a gauge where $A_5=0$, we perform a KK decomposition of the field, similar to the scalar case in eq.~$(\ref{KKscalar})$. We impose Neumann boundary conditions and  canonical normalisation of the 4D kinetic terms to find the KK profiles, which we denote by $w_n(y)$.  We will account for EWSB masses in the 4D theory therefore our zero mode will be flat and massless, hence,
\begin{equation} \label{gaugezero}
w_0(y)=1
\end{equation}
and the tower of massive KK modes will have profiles described by \cite{Gherghetta:2000kr}
\begin{equation}
w_n(y)=\frac{e^{ky}}{N_n}\left(J_{1}\left(\frac{{m}_n}{k}e^{ky}\right)+\beta(m_n)Y_{1}\left(\frac{{m}_n}{k}e^{ky}\right)\right).
\end{equation}
Here the $m_n$ are the gauge boson KK masses. In the limit $kL>>1$ the KK masses can be approximated as
\begin{equation}
m_n\simeq \left(n-\frac{1}{4}\right)\pi k e^{-kL} .
\end{equation}
Due to the KK masses, any local gauge symmetries in the bulk only survive as global symmetries of the KK spectrum in the effective theory.  It is only the massless modes in the spectrum that remained gauged, thus the global symmetry is said to be ``weakly gauged".

\subsection{Fermion fields}
The treatment of fermions is complicated slightly by the fact that 5D Dirac fermions are not chiral.  For a detailed discussion of fermions in 5D see \cite{Csaki:2005vy,Gherghetta:2010cj}.  We will refrain from such a discussion here.  The action and resulting equation of motion for the 4-component 5D Dirac fermion can be written as
\begin{equation}
\begin{gathered}
S_{\Phi}=\int d^4x\int_0^Ldy\sqrt{|g|}\left(\frac{1}{2}\left(\bar{\Psi}\gamma^MD_M\Psi-D_M\bar{\Psi}\gamma^M\Psi\right)-m_{\Psi}\bar{\Psi}\Psi\right), \\\\
E_a^M\gamma^a\left(\partial_M+\omega_M\right)\Psi -m_{\Psi}\Psi=0,
\end{gathered}
\end{equation}
where $E_a^M$ is the f\"{u}nfbein, $E_a^M\gamma^a=\gamma^{M}$, $\gamma^a=(\gamma^{\mu},i\gamma^5)$ are the gamma matrices in flat space, and $\omega_M$ is the spin connection \cite{Grossman:1999ra,Gherghetta:2000kr}.  We can write the 5D fermion as 
\begin{equation}
\Psi=\Psi_++\Psi_-=\left(\begin{array}{c} \psi_+ \\ 0 \end{array}\right)+\left(\begin{array}{c} 0 \\ \psi_- \end{array}\right),
\end{equation}
where $\Psi_{\pm}=\pm\gamma_5\Psi_{\pm}$ denotes left and right handed components, respectively.  We can then write a second order equation of motion for these fields as
\begin{equation}
\left(e^{2ky}\partial_{\mu}\partial^{\mu}+\partial_y^2-k\partial_y-c(c\pm1)k^2\pm ck(\delta(y)-\delta(y-L))\right)\Psi_{\pm}=0,
\end{equation}
where we have written the 5D fermion mass term as $m_{\Psi}={\rm sign}(y)ck$.  The non-trivial boundary terms arising here allow us to have localised fermion zero modes.  Performing a KK decomposition on the field, it has been shown that by applying Dirichlet boundary conditions to $\Psi_+$ ($\Psi_-$) ensures that the zero mode will be right (left) handed.  Thus from our choice of boundary conditions we ensure that for each Weyl fermion in the SM, the 4D effective theory will only contain the corresponding zero mode Weyl fermion plus a tower of vector-like KK Dirac fermions.  One can then arrive at the following 5D profile for the massless zero mode \cite{Grossman:1999ra,Gherghetta:2000kr}
\begin{equation} \label{fermionzero}
f_{\pm}^{(0)}(y)=\frac{1}{N^{(0)}_{\pm}}e^{(2\mp c)ky}
\end{equation}
and can also obtain the profiles of the vector-like KK states of these fields as
\begin{equation} \label{fermionKK}
f_{\pm}^{(n)}(y)=\frac{e^{ky/2}}{N^{(n)}_{\pm}}\left[J_{c\pm\frac{1}{2}}\left(\frac{m_n}{k}e^{ky}\right)+\beta(m_n)Y_{c\pm\frac{1}{2}}\left(\frac{m_n}{k}e^{ky}\right)\right].
\end{equation}
The fermionic KK masses are denoted by $m_n$. By varying the 5D mass parameter $c$ we can localise the zero modes anywhere in the bulk, whereas the KK modes will always be IR localised \cite{Grossman:1999ra,Gherghetta:2000kr,Huber:2000ie}.  In the same fashion as before we have that in the limit $kL>>1$ the KK masses can be approximated as
\begin{equation}
m_n\simeq \left(n+\frac{|\alpha|}{2}-\frac{1}{4}\right)\pi k e^{-kL}.
\end{equation}
The spectrum of KK masses and the constants $\beta(m_n)$ depend on which zero mode chirality we have chosen.  In expressions (\ref{fermionKK}) and (26) we have $\alpha=c\pm\frac{1}{2}$ for $\Psi_{\pm}$ obeying Dirichlet boundary conditions, i.e. a right (left) handed zero mode implies $\alpha=c+\frac{1}{2}$ ($\alpha=c-\frac{1}{2}$).  Note that the masses and profiles of the KK modes of the right handed zero mode field are generally different than those of the left handed zero mode field, given a fixed value of $c$.

\section{The Higgs potential in RS}
The model we now wish to study consists of an SU($2$) Higgs doublet $\Phi$ of complex scalars living in a slice of AdS space.  The 5D action for this system is
\begin{equation}
S=\int d^4x \int_0^{L} dy \hspace{1mm}e^{-4ky}\left( (D^{M}\Phi)^{\dagger}(D_{M}\Phi)-m_{\Phi}^2\Phi^{\dagger}\Phi-\lambda_{5} (\Phi^{\dagger}\Phi)^2\right)
\end{equation}
and
\begin{equation}
\Phi(x^{\mu},y)=\left( \begin{array}{c}
\phi^+(x^{\mu},y) \\
\phi^0(x^{\mu},y) \\
\end{array} \right),
\end{equation}
where $\phi^+$ and $\phi^0$ are complex scalar fields.  The mass term, defined in eq.~(\ref{scalarmass}), and quartic coupling in our model can have localised brane contributions and in principle can be $y$-dependent in the bulk (however we assume them to be constant).  The quartic coupling then is of the form
\begin{equation}
\lambda_5=\lambda_B+\frac{1}{k}\lambda_{IR}\delta(y-L)+\frac{1}{k}\lambda_{UV}\delta(y).
\end{equation}
The $\lambda_{UV}$ term is irrelevant for an IR scalar, here thus we will only consider the IR contribution.  In this section we will study models in which we have a quartic term on the brane and/or in the bulk.  In both cases we go to the 4D theory before we treat the breaking of SU($2$).  We will then comment on the relation between these effective theories and theories involving multiple Higgs doublet models.

\subsection{Brane EWSB}
This case is straight forward.  To reach the 4D effective theory we follow a method exactly like that in section 2.1 and find the same scalar profiles.  The only difference is an extra term in the effective action corresponding to the brane quartic coupling
\begin{equation}
S=\int d^4x \frac{1}{2} \left( \sum_n|\partial_{\mu}\Phi_n|^2-m_n^2\Phi^{\dagger}_n\Phi_n-\sum_{m,n}\delta m_{mn}^2\Phi^{\dagger}_m\Phi_n-\lambda_{lmnp}\sum_{lmnp}\Phi^{\dagger}_l\Phi_m\Phi^{\dagger}_n\Phi_p\right),
\end{equation}
where
\begin{equation}
\lambda_{lmnp}=\frac{\lambda_5}{L}e^{-4kL}f_l(L)f_m(L)f_n(L)f_p(L)
\end{equation}
and $\delta m_{mn}^2$ is defined in eq.~(\ref{scalarmasscorr}).  The standard model Higgs will be identified with the lightest mass eigenstate, being predominately composed of $\Phi_0$.
Taking the approximation with just the  zero mode plus first $N$ KK states, we have $(N+1)$ Higgs doublets in our effective theory. From here we can minimise the potential and find expressions for the vacuum expectation values of these fields, $\langle \Phi_m\rangle=v_m$.  
The largest correction to the standard model Higgs potential will be of the form $\lambda_{1000}\Phi_0^3\Phi_1$, making $\lambda_{1000}$ the most important BSM coupling in this sector.  

\subsection{Bulk EWSB}
We write the scalar doublet so that we can see clearly the excitations around its minimum,
\begin{equation}
\Phi(x^{\mu},y)= \frac{1}{\sqrt{2}}\left( \begin{array}{c}
 \phi^{+}(x^{\mu},y)\\
v(y)+\phi^{0}(x^{\mu},y) \\
\end{array} \right).
\end{equation}
With a quartic term in the bulk we can write the total energy functional of the 5D system in the ground state as
\begin{equation}
E[v(y)]=\int dx^3\int_0^{L}dy\frac{1}{2}\sqrt{|g|}\left( (\partial_{y}v)^2+m_{\Phi}^2v^2+\lambda_5 v^4 \right).
\end{equation}
Minimising this, we find that in the ground state the field must obey the following EOM
\begin{equation}
-\frac{1}{\sqrt{|g|}}\partial_y(\sqrt{|g|}\partial_yv)+b^2k^2v+\lambda_B v^3=0.
\end{equation}
Boundary term in the scalar mass  will induce non-trivial boundary condition, similar to the discussion in section 2.1.
We choose a gauge in which we can write the doublet as
\begin{equation}
\Phi(x^{\mu},y)= \frac{1}{\sqrt{2}}\left( \begin{array}{c}
 0\\
v(y)+\eta(x^{\mu},y) \\
\end{array} \right),
\end{equation}
where $\eta={\rm Re}(\phi^{0})$.  We can now write the action for the physical field $\eta$ as
\begin{equation}
\begin{gathered}
S=\int d^4x \int_0^{L} dy \frac{1}{2}\sqrt{|g|} \left(e^{2A}\frac{1}{2}\partial^{\mu}\eta\partial_{\mu}\eta - \frac{1}{2}\left(-\frac{1}{\sqrt{|g|}}\partial_y(\sqrt{|g|}\partial_y\eta)+b^2k^2\eta+\lambda_B v^2\eta\right)\eta \right. \\ \left.
-\frac{\lambda_B}{4}\eta^4 - \lambda_B v\eta^3 + \lambda_B v^4 \right),
\end{gathered}
\end{equation}
where $A(y)=k|y|$ denotes the warp factor.
Expanding $\eta$ into KK modes to diagonalize the fields in the mass eigenbasis, the equation of motion for the 5D profiles reads
\begin{equation}
-\frac{1}{\sqrt{|g|}}\partial_y(\sqrt{|g|}\partial_yf_n)+b^2k^2\eta+\lambda_B v^2f_n=\sqrt{|g|}e^{2A}m_n^2f_n.
\end{equation}
Again nontrivial boundary conditions are induced by the brane masses. Here $m_n^2$ is the mass of the $n$th KK mode. Thus $m_0^2$ and $f_0$ refer to the physical Higgs mode.  This shows us that the Higgs and the vacuum expectation value have different 5D profiles, thus their interaction with the fermion and gauge fields will differ from the standard model.  This difference is determined by the Higgs mass and the KK scale \cite{Azatov:2009na}.

Ideally we would like to solve these non-linear differential equations and have the correct 5D profiles for the mass eigenbasis at our disposal. However, it is difficult to obtain reliable numerical solutions.  Instead we will not diagonalize the fields in the mass eigenbasis, but will expand them in the basis (\ref{scalaraction}) we used in section 2.1.  Hence we use the same 5D profile for the zero mode and the vacuum expectation value.  This will result in an effective theory similar to that in the brane EWSB case, except now the effective quartic term is given by
\begin{equation} \label{quartics}
\lambda_{lmnp}=\frac{\lambda_5}{L^2}\int_0^Ldy\sqrt{|g|}f_lf_mf_nf_p.
\end{equation}
The only difference we have is that the relationship between the different quartic couplings changes.  In table \ref{quarticstable} we show the values of these bulk and brane quartics for $a^2=-2$ and take the two cases $\lambda_B=1$ and $\lambda_{IR}=1/4$.  The effects of KK modes in the Higgs sector are usually proportional to the quartic couplings, the largest effect is $\sim \lambda_{1000}v_1$ and hence the most relevant coupling is $\lambda_{1000}$.  From table 1 we see that having bulk EWSB terms reduces the higher mode quartic terms with respect to $\lambda_{0000}$ and will therefore reduce the KK effects in general.

\begin{table}[t]
\centering
    \begin{tabular}{lccccc}
    &${\lambda}_{0000}$ & $\lambda_{1000}$ & $\lambda_{1100}$ &$\lambda_{1110}$ & $\lambda_{1111}$  \\ \hline
     Brane Quartic&1.00 &-1.00&1.00&-1.00&1.00 \\
     Bulk Quartic&1.00 &-0.54&0.66&-0.34&0.70 \\
\end{tabular}
\caption{This shows the values of the quartic couplings for brane and bulk EWSB with $a^2=-2$ and $\lambda_B=1$ or $\lambda_{IR}=1/4$.}
\label{quarticstable}
\end{table}

The SM particles receive small mass corrections from the v.e.v. of KK Higgs fields, which indices a misalignment of Higgs couplings and particle masses.  We find that KK v.e.v.'s are approximately
\begin{equation} \label{vevs}
v_n\simeq -\frac{\lambda_{n000}}{\lambda_{0000}}\frac{m_H^2}{m_n^2}v_0.
\end{equation} 
From table 1 we see that the ratio for a brane quartic coupling is $\lambda_{1000}/\lambda_{0000}=-1$, and for a bulk quartic $\simeq -0.5$.  
In the next section we will see that electroweak constraints force KK resonances into the multi-TeV range, thus leading to $v_n/v_0$ in the  sub per-mille range. The resulting impact on couplings between the Higgs and gauge bosons is then also in the sub per-mille range, and too small to make an impact even at TLEP \cite{Davoudiasl:2005uu}.  (A different and potentially observable source of modifications of the gauge Higgs coulings we will discuss in the the next section.)
Also modifications of the Higgs cubic self coupling are at similar level and thus too small to be observed.
The other important factor is the coupling of SM particles to the Higgs KK modes.  Gauge zero modes have flat profiles, hence the normalisation of the Higgs field ensures that they couple equally to all Higgs KK modes.  For the fermion fields we also find that the fermion zero modes couple to Higgs zero and KK modes almost identically, as the Higgs is IR localised.

\subsection{Multiple Higgs Doublet Models}
Since the tower of Higgs doublets in the effective theory all couple to the up and down type quarks, this could be viewed as a theory of multiple Higgs doublets with v.e.v.'s given by eq.~(\ref{vevs}) and couplings given by eq.~(\ref{quartics}).  If we include one additional mode for simplicity, we have a type III 2HDM which is well studied phenomenologically.  The experimental constraints for these models are summarised in \cite{Atlas2HDM}. They express the constraints in terms of $\tan(\beta)=v_1/v_0$ and $\cos(\beta-\alpha)$, the ratio of the Higgs KK mode and zero mode couplings to the SM gauge bosons.  In our model both these observables are $\sim \frac{v^2}{M_{KK}^2}$, i.e.~per-mille, and well within the experimental constraints.  For these bounds to be relevant we would need $M_{KK}\lesssim1$ TeV.

\section{Electroweak precision observables}
\subsection{Calculable Corrections}
We consider a non-custodial $SU(2)_L\times U(1)_Y$ bulk gauge sector with bulk fermions and a bulk Higgs.  Calculating the Peskin-Takeuchi parameters \cite{Peskin:1991sw} is straightforward, and assuming universal UV fermion localisations for the light fermions, we can account for their effects also.  Corrections to the SM can arise from the zero mode fields mixing with KK modes, and from the exchange of KK particles in a physical process. For our purposes the latter is only a small effect and will be ignored.  For a detailed analysis of the case of a brane Higgs see e.g.~\cite{Delgado:2007ne}.  Our low energy 4D effective theory can be written in the form \cite{Csaki:2005vy}
\begin{equation} \label{EWlag}
\begin{gathered}
\mathcal{L}=-\frac{1}{4}F^{\mu\nu}F_{\mu\nu}-\frac{1}{2}W^{\mu\nu}W_{\mu\nu}-\frac{1}{4}Z^{\mu\nu}Z_{\mu\nu} - \frac{1}{2}(1+\delta z)m_Z^2Z^{\mu}Z_{\mu} -(1+\delta w)m_W^2W^{\mu}W_{\mu} \\
-e(1+\delta a^{\psi})\sum_i \bar{\psi}_i\gamma^{\mu}Q_i\psi_iA_{\mu} - \frac{e}{s_W\sqrt{2}}(1+\delta w^{\psi})\sum_{ij}(V_{ij}\bar{\psi}_i\gamma^{\mu}P_L\psi_jW_{\mu}^+ + c.c. ) \\
- \frac{e}{s_Wc_W}(1+\delta z^{\psi})\sum_i \bar{\psi}_i\gamma^{\mu}\left[T_{3i}P_L-Q_is_W^2+Q_is_Wc_W\lambda_{ZA}\right]\psi_iZ_{\mu},
\end{gathered}
\end{equation}
where $V_{ij}$ is the CKM matrix and $\delta z$, $\delta w$, $\delta a^{\psi}$, $\delta w^{\psi}$ and $\delta z^{\psi}$ are flavour independent new physics contributions to the Lagrangian.  From this Lagrangian we can identify the $S$, $T$ and $U$ parameters with
\begin{equation}
\begin{gathered}
\alpha S=4s_W^2c_W^2(-2\delta a^{\psi}+2\delta z^{\psi}) \\
\alpha T=(\delta w-\delta z)-2(\delta w^{\psi}-\delta z^{\psi}) \\
\alpha U=8s_W^2(-\delta a^{\psi}s_W^2+\delta w^{\psi}-c_W^2\delta z^{\psi}).
\end{gathered}
\end{equation}
We decompose the 5D $SU(2)_L$ and $U(1)_Y$ gauge fields as
\begin{equation}
W^{M 3}=\frac{1}{\sqrt{L}}\sum_n w_n(y)W_n^{\mu 3}(x^{\mu}), \hspace{2mm}  W^{M \pm}=\frac{1}{\sqrt{L}}\sum_n w_n(y)W_n^{\mu \pm}(x^{\mu}), \hspace{2mm} B^{M}=\frac{1}{\sqrt{L}}\sum_n w_n(y)B_n^{\mu}(x^{\mu}), 
\end{equation}
where $w_0=1$ as defined in eq.~(\ref{gaugezero}).  In unitary gauge the Higgs can be written in the following form
\begin{equation}
\Phi(x^{\mu},y)=\frac{1}{\sqrt{2L}}f_0(y)\left( \begin{array}{c} 0 \\ v_0 + h(x^{\mu}) \end{array} \right),
\end{equation}
where $f_0$ is given from eq.~(\ref{scalarzero}) and we ignore KK Higgs modes.  When we go to the 4D effective theory, we can write the mass matrices for the gauge fields  as
\begin{equation}\label{WmassM}
M_W^2=\frac{g^2}{4}\left(\begin{array}{ccc}
M_{00}^2  & M_{01}^2 &\hdots \\
M_{01}^2 & \frac{4}{g^2}m_1^2+M_{11}^2 & \hdots \\
\vdots & \vdots & \ddots \\
\end{array} \right)
\end{equation}
\begin{equation}
M_Z^2=\frac{g^2+g^{\prime 2}}{4}\left(\begin{array}{ccc}
M_{00}^2  & M_{01}^2 & \hdots \\
M_{01}^2 & \frac{4}{g^2+g^{\prime 2}}m_1^2+M_{11}^2 & \hdots\\
\vdots & \vdots & \ddots \\
\end{array} \right)
\end{equation}
\begin{equation}
M_{\gamma}^2=\left(\begin{array}{ccc}
0  & 0 & \hdots \\
0 & m_1^2 & \hdots \\
\vdots & \vdots & \ddots \\
\end{array} \right),
\end{equation}
where
\begin{equation}
M_{mn}^2=\frac{v_0^2}{L}\int_0^Ldy\hspace{1mm}e^{-2ky}w_mw_nf_0^2
\end{equation}
and $m_n^2$ are the gauge KK masses.   The normalisation of the Higgs field means that $M_{00}^2=v^2_0$.  We can approximately diagonalize these mass matrices assuming that $M_{00}^2,M_{0n}^2<<m_n^2$, and find lowest mass eigenvalues to be
\begin{equation} \label{WZmass}
\begin{gathered}
\left(M_W^2\right)_{0}\simeq \frac{g^2v_0^2}{4}\left(1 - \frac{g^2v_0^2}{4}\sum_n \frac{R_n^2}{m_{n}^2}\right) \\
\left(M_Z^2\right)_{0}\simeq \frac{(g^2+g^{\prime 2})v_0^2}{4}\left(1 - \frac{(g^2+g^{\prime 2})v_0^2}{4}\sum_n \frac{R_n^2}{m_{n}^2}\right), 
\end{gathered}
\end{equation}
where $R_n={M_{0n}^2}/{v_0^2}$ and parametrizes the coupling between the Higgs and gauge excitations. The photon remains massless.  In moving to the mass eigenbasis  the fermion and Higgs couplings to the W and Z bosons get shifted.  We are only interested in the shift in the fermion-gauge coupling since, at tree-level, the gauge-Higgs couplings do not alter the electroweak precision analysis.  We write the unshifted vertex term between a fermion and the W boson as
\begin{equation}
\sum_n\frac{g_{0n}}{\sqrt{2}s_W}\sum_{i}(V_{i0}\bar{\psi}_{i0}\gamma^{\mu}P_L\psi_{j0}W_{\mu n}^+ + c.c. ),
\end{equation}
where $g_{mn}$ is the effective coupling,
\begin{equation}
g_{mn}=\frac{g_5}{L^{\frac{3}{2}}}\int_0^Ldy\hspace{1mm}e^{-3ky}(f_{+}^{(m)})^2w_n
\end{equation}
and $f_+^{m}$ is defined in eqs.~(\ref{fermionzero}) and (\ref{fermionKK}).  When we go to the mass eigenbasis, the interaction of the fermion with the zero mode gauge field is of the form
\begin{equation}
\frac{g_{00}}{\sqrt{2}{s}_W}\left(1-\frac{g^2}{4}\sum_n\frac{M_{0n}^2}{m_n^2}\frac{g_{0n}}{g_{00}}\right)\sum_{i}(V_{ij}\bar{\psi}_{i0}\gamma^{\mu}P_L\psi_{j0}W_{\mu 0}^+ + c.c. ).
\end{equation}
For the Z coupling we have an analogous expression proportional to $g^2+g^{\prime 2}$.  Since the photon remains massless the photon vertices do not get extra contributions.  With this information we can express the electroweak parameters as
\begin{equation} \label{ST}
\begin{gathered}
S\simeq\left(\frac{-9\pi}{2}\sum_n\frac{R_n}{(n-\frac{1}{4})^2}\frac{g_{0n}}{g_{00}}\right)\frac{v_0^2}{M_{KK}^2} \\
T\simeq\left(\frac{9\pi}{16c_W^2}\sum_n\frac{R_n}{(n-\frac{1}{4})^2}\left(R_n+2\frac{g_{0n}}{g_{00}}\right)\right)\frac{v_0^2}{M_{KK}^2} \\
\end{gathered}
\end{equation}
whereas $U\sim\left(g^2-(g^2+g^{\prime 2})c_W^2\right)=0$.  In the above calculation we used the expressions for the gauge KK masses in section 2.2 and have taken $M_{KK}=m_1\simeq(3\pi/4)ke^{-kL}$, i.e.~the mass of the first gauge boson excitation.  From the expressions in eq.~(\ref{ST}), neglecting contributions from higher KK modes, we find a correlation between the $S$ and $T$ parameters which can be expressed as
\begin{equation}\label{STcorr}
T\simeq\frac{1}{8c_W^2}\left(2-\frac{g_{00}}{g_{01}}R_1\right)S.
\end{equation}
Depending on $T/S$, the model can live in more or less experimentally favoured regions of the parameter space, possibly resulting in reductions to the $M_{KK}$ constraint.   

To a good approximation $R_n$ and $g_{0n}/g_{00}$ do not vary with $L$, meaning that the only $L$ dependence in $S$ and $T$ comes from $M_{KK}$.  $R_n$ varies with the Higgs localisation, and becomes smaller as the Higgs leaks to the bulk.  Table \ref{higgstable} shows that the bulk Higgs couples less to gauge KK modes than the brane Higgs.  As a result, not only will the $T$ parameter be smaller for a bulk Higgs, but we find that a two mode approximation is accurate for a bulk Higgs, but not sufficient for a brane Higgs.

\begin{table}[t]
\centering
\begin{tabular}{lllll}
 & $R_1$ & $R_2$ & $R_3$ & $R_4$ \\
Brane Higgs & 8.4 & -8.3 & 8.1 & -8.2 \\
Bulk Higgs ($a^2=-2$) & 5.6 & -0.9 & 0.5 & -0.3 
\end{tabular}
\caption{Here we show how the couplings between the zero mode Higgs and the gauge KK tower differ for a brane and bulk Higgs.}
\label{higgstable}
\end{table}

\begin{figure}[t] 
\centerline{\includegraphics[scale=0.18]{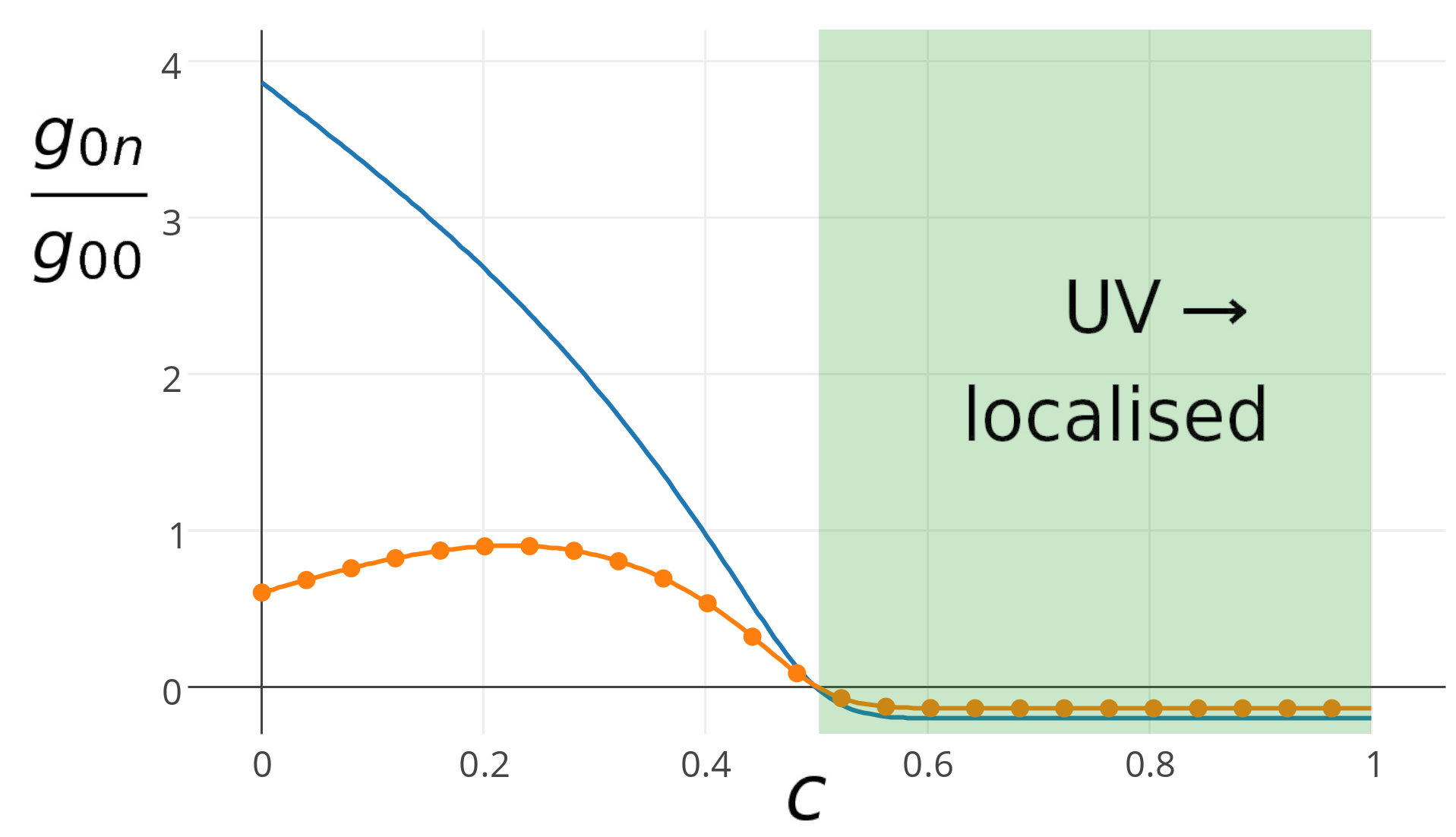}}
\caption{This plot shows ${g_{0n}}/{g_{00}}$ over a range of fermion localisations for $n=1$ (solid) and $n=2$ (dotted).  The shaded region shows the parameter space for which the fermions are UV localised.}
\label{fermiongraph}
\end{figure}
Light fermions must be localised in the UV so that their overlap with the Higgs is small, this corresponds to $c_L>0.5$ \cite{Grossman:1999ra,Gherghetta:2000kr,Huber:2003tu}.  From figure \ref{fermiongraph} we can see that this implies a small coupling with the KK gauge modes and therefore small vertex contributions to the electroweak parameters.  For all fermion localisations we find that the coupling decreases for heavier KK modes.
\newline
\newline
Current bounds on $S$ and $T$ with $U=0$ are given in \cite{Baak:2012kk} (see figure \ref{figurest}). Taking the 95\% CL bound, we find the following bounds for a brane and bulk Higgs:
\begin{itemize}
\item \underline{Brane Higgs}:  Due to the large values of $R_n$ the KK gauge modes have large contributions to the $T$ parameter.  If we approximate $R_n^2\simeq8.4^2$ for all $n$, we can sum the full tower contributions by taking the sum $\sum_{n=1}^{\infty}(n-1/4)^{-2}\simeq2.54$.  We then find that the electroweak constraints require $M_{KK}\gtrsim15$ TeV.
\item \underline{Bulk Higgs} ($a^2=-2$):  Since the $R_n$ values are small for $n>1$ we find that the first mode makes the only sizeable contribution to the electroweak parameters.  With just the first excited mode we find the bounds to be $M_{KK}\gtrsim8$ TeV.  Including the first 10 modes only corrects the $8$ TeV bound by $0.3\%$, and the second excited mode contributes $0.26\%$ of this correction.  We find similar effects for the $S$ parameter.
\end{itemize}
These results are in agreement with the bounds found elsewhere in the literature \cite{Cabrer:2011fb,Fichet:2013ola,Archer:2014jca}.

Another thing one should consider is the misalignment in the gauge boson masses and their coupling to the Higgs zero mode.  The couplings between the Higgs zero mode and the gauge modes can be written as a matrix similar to eq.~(\ref{WmassM}) but without the large contributions from the KK masses.  The absence of the KK masses here is what causes the misalignment when we go to the mass eigenbasis.  We find that the HHZ and the HHZZ interactions receive identical corrections $\sim R_1^2\hspace{1mm}m_Z^2/M_{KK}^2$,  and similarly for the W boson.  With the lightest gauge boson mass at $8$ TeV we find a $0.4\%$ misalignment for the Z boson and a $0.3\%$ misalignment for the W boson.  This would be visible at the ILC \cite{Peskin:2012we,Asner:2013psa} or TLEP \cite{Gomez-Ceballos:2013zzn}.  The only way to reduce this misalignment is to either increase $M_{KK}$ or to reduce the coupling of the Higgs zero mode to the gauge KK modes, which can be achieved by modifying the background geometry in the bulk \cite{Falkowski:2008fz,Cabrer:2010si,Cabrer:2011fb,Cabrer:2011mw,Carmona:2011ib}.

\subsection{Higher dimensional operator contributions to $S$, $T$ and $U$}

In the previous section we demonstrated how to estimate the size of the calculable contributions to the electroweak parameters in the 4D effective theory.  There will also be incalculable contributions from the UV theory which we will parameterise using higher dimensional operators in the 5D theory.  The three leading operators contributing to the oblique parameters are
\begin{align}
S: \hspace{3mm}&\frac{\rho}{M_5^3}(\Phi^{\dagger}T^a\Phi)W^a_{MN}B_{MN} \nonumber \\
T:  \hspace{3mm}&\frac{\lambda}{M_5^3}|\Phi^{\dagger}D^M\Phi|^2 \nonumber \\
U: \hspace{3mm}&\frac{\theta}{M_5^6}|\Phi^{\dagger}W^{MN}\Phi|^2,
\end{align}
where $\rho$, $\lambda$ and $\theta$ are unknown parameters. These operators could be present both on the branes or in the bulk, i.e.~$\rho=\rho_B+\rho_{IR}M_5^{-1}\delta(y-\pi R)$.  In the brane case there is an extra mass scale suppression. There is also a possible contribution from the UV brane, which is negligible for an IR localised Higgs.

The $S$ and $T$ operators both have effective coefficients $\sim v_0^2/M_{KK}^2$, but due to the higher dimension of the $U$ operator it is of the order $\sim \left(v_0^2/M_{KK}^2\right)^2$.  Thus only $S$ and $T$ will receive sizeable corrections from these operators, while $S$ also has an additional suppression $\sim \frac{1}{kL}$ with the respect to the $T$ coefficient.  All three operators show similar dependence on the Higgs localisation. The effective coefficients grow exponentially as $a^2$ decreases until, at $a^2=-1$, the exponential growth stops, which is due to the normalisation of the Higgs field.  At $a^2<-1$, operators on the IR brane increase linearly with $a^2$ while operators in the bulk remain mostly constant.  At $a^2=-2$ the operator coefficients from the IR brane contributions are
\begin{align}
\rho_{IR} &\rightarrow \alpha\delta S=\rho_{IR}(kL)^{-1}\kappa^4\left(\frac{v_0}{ke^{-kL}}\right)^2 \nonumber \\
\lambda_{IR}&\rightarrow  \alpha\delta T=-4\lambda_{IR}\kappa^4\left(\frac{v_0}{ke^{-kL}}\right)^2 \nonumber \\
\theta_{IR}&\rightarrow \alpha\delta U=-4\theta_{IR}(kL)^{-1}\kappa^7\left(\frac{v_0}{ke^{-kL}}\right)^4
\end{align}
and in the bulk are
\begin{figure}[t]
 \centerline{\includegraphics[scale=0.13]{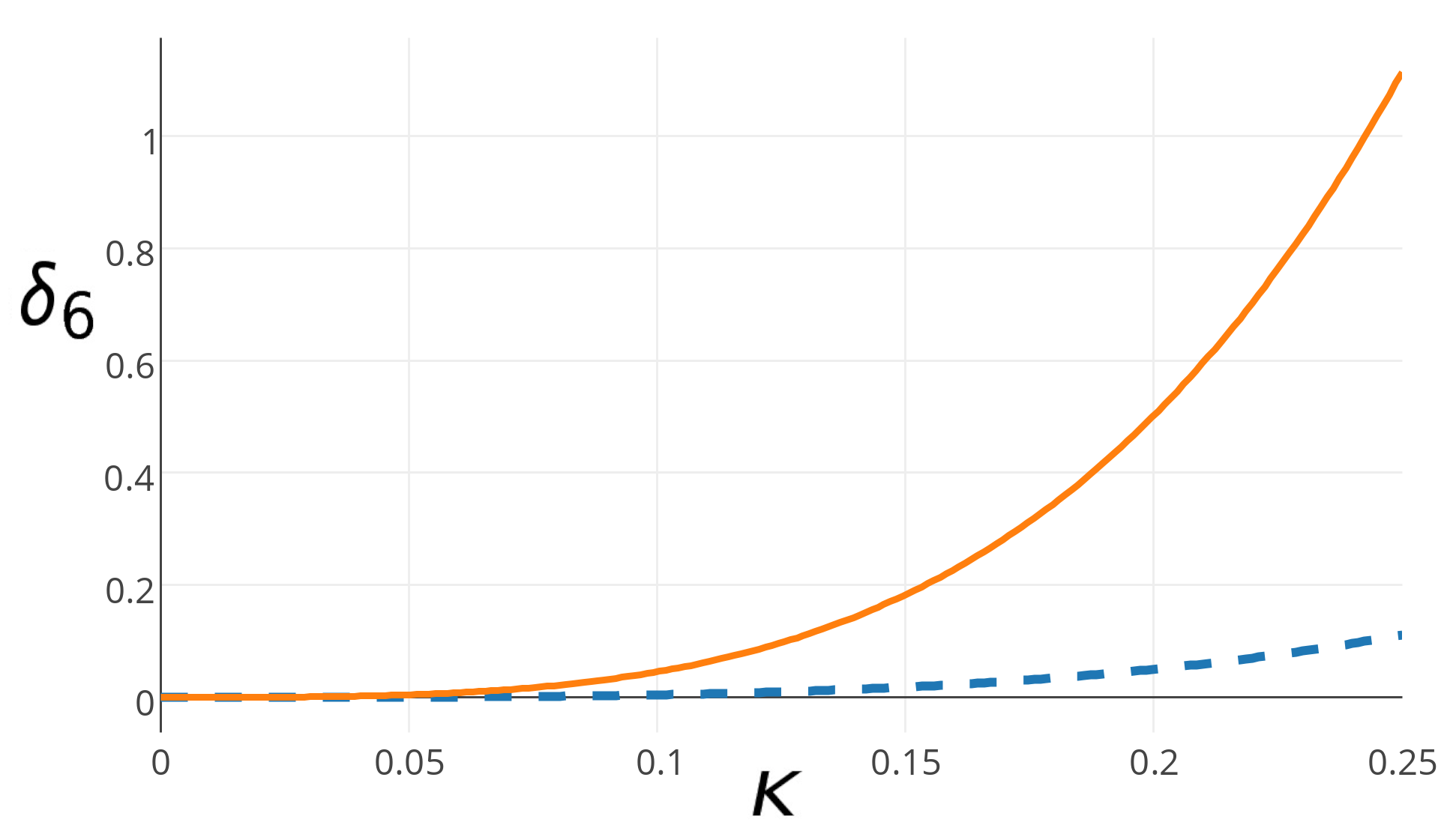}}
  \caption{Here we show how $\delta_6$ varies with $\kappa$ for $\lambda_B=\lambda_{IR}=1$ (dashed) and $\lambda_B=\lambda_{IR}=10$ (solid).}
  \label{figuredelta8}
\end{figure}
\begin{align}
\rho_{B}&\rightarrow \alpha\delta S=\rho_B(2kL)^{-1}\kappa^3\left(\frac{v_0}{ke^{-kL}}\right)^2 \nonumber \\
\lambda_{B}&\rightarrow\alpha\delta T= -\frac{2}{3}\lambda_B\kappa^3\left(\frac{v_0}{ke^{-kL}}\right)^2 \nonumber \\
\theta_{B}&\rightarrow \alpha\delta U=-\theta_B(2kL)^{-1}\kappa^6\left(\frac{v_0}{ke^{-kL}}\right)^4,
\end{align}
where the $B$ and $IR$ subscripts refer to the bulk and brane parameters, respectively.  Again we set $\kappa=k/M_5$. With $\mathcal{O}(1)$ values for the operator coefficients we would only expect a sizeable contribution from the operator contributing to the $T$-parameter.  With respect to this operator, the operator contributing to the $U$ parameter is suppressed by two additional powers of mass, and the operator contributing to the $S$ parameter has an additional volume suppression.  This behaviour in the $U$ parameter has been noted in \cite{Barbieri:2004qk} also.  If we ignore the vertex corrections, and include the effects of these operators in our $T$ parameter expression from eq.~(\ref{ST}), we find total $T$ parameter
\begin{equation}
T_6\simeq\left(\frac{3\pi}{4}\right)^2\left(\frac{1}{\pi c_W^2}\sum_n\frac{R_n^2}{(n-0.25)^2}+\frac{2}{3}\frac{\kappa^3}{\alpha}\lambda_B+4\frac{\kappa^4}{\alpha}\lambda_{IR}\right)\frac{v_0^2}{M_{KK}^2}=T\left(1+\delta_6\right),
\end{equation}
where we again took $M_{KK}=m_1\simeq(3\pi/4)ke^{-kL}$.
Here $\delta_6$ parameterises the contribution from higher dimensional operators,
\begin{equation}
\delta_6=\left(\frac{1}{\pi c_W^2}\sum_n\frac{R_n^2}{(n-0.25)^2}\right)^{-1}\left(\frac{2}{3}\frac{\kappa^3}{\alpha}\lambda_B+4\frac{\kappa^4}{\alpha}\lambda_{IR}\right).
\end{equation}
From figure \ref{figuredelta8} we see it may be reasonable to argue that these contributions could be large enough to provide a reduction in the $T$ parameter calculated in eq.~(\ref{ST}).  This also modifies eq.~(\ref{STcorr}) such that the correlation is expressed as
\begin{equation}
T_6\simeq\frac{1}{8c_W^2}\left(2-\frac{g_{00}}{g_{01}}R_1\right)\left(1+\delta_6\right)S.
\end{equation}
From figure \ref{figurest} we see that as well as directly reducing the $T$ parameter, $\delta_6\neq0$ can take us to a more favourable region of the parameter space, depending on the relative sign, thus allowing for a further reduction on the $M_{KK}$ bound.  If we take the 95\% CL bound from figure 4, we find that the lower bound on $M_{KK}$ is approximately $6$ TeV and $2.7$ TeV for $\delta_6=-0.4$ and $-0.8$, respectively. So it is plausible to assume that incalculable contributions to the $T$ parameter lead to a partial cancellation and so relax the bound on the KK scale. It therefore seems premature to exclude discovery of such a scenario at the forthcoming LHC run.

\begin{figure}[t]
\centerline{\includegraphics[scale=0.25]{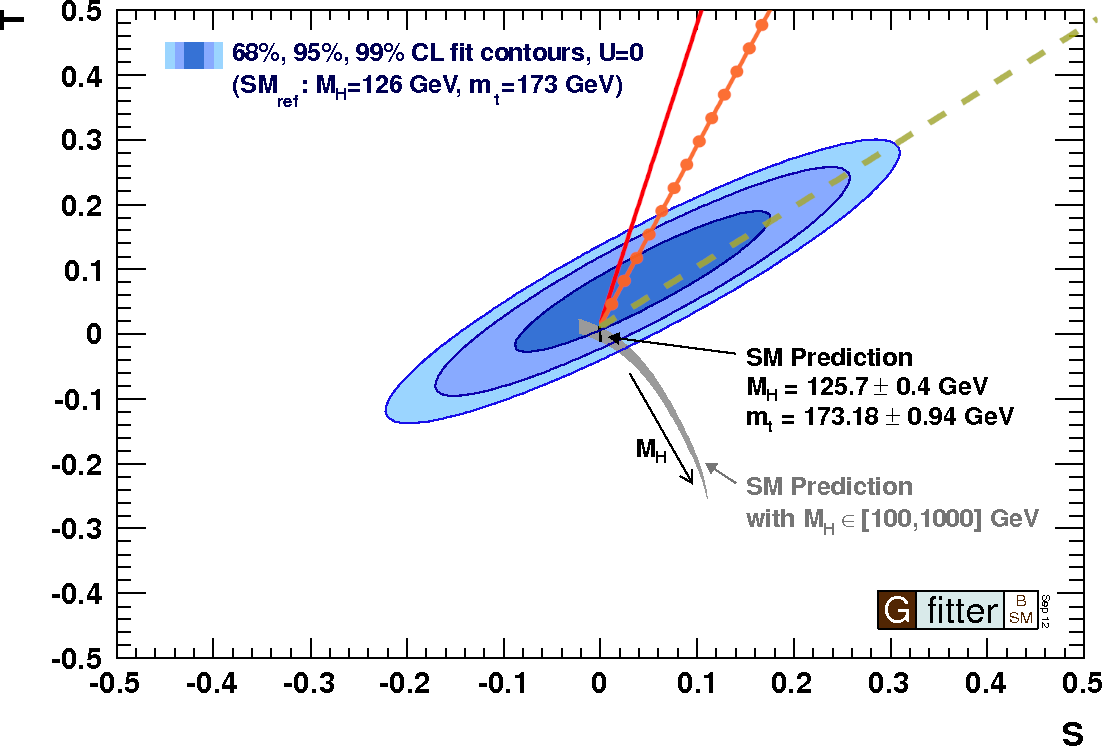}}
\caption{Here we have overlaid the bounds from \cite{Baak:2012kk} with the $S$ and $T$ correlations for $\delta_6=0$ (solid), $\delta_6=-0.4$ (dots) and $\delta_6=-0.8$ (dashed).}
\label{figurest}
\end{figure}

\section{Yukawa coupling corrections}
The aim of this section is to investigate possible bounds on the bulk Higgs scenario from corrections to SM Yukawa couplings. 
Consider an SU(2) singlet fermion $t$ and doublet $Q=(T,B)$ in the 5D theory. The action for such a system, omitting terms in $B$, can be written as \cite{Grossman:1999ra,Gherghetta:2000kr}
\begin{equation}
\begin{gathered}
S=\int d^4x \int_0^L dy\hspace{1mm}\sqrt{|g|}\left(\frac{1}{2}\left(\bar{t}\gamma^MD_Mt-D_M\bar{t}\gamma^Mt\right)-m_t\bar{t}t \right. \\ \left. 
+\frac{1}{2}\left(\bar{T}\gamma^MD_MT-D_M\bar{T}\gamma^MT\right)-m_T\bar{T}T
+\lambda_t^{(5)}\sqrt{L}\phi^0\bar Tt + {\rm h.c.}\right),
\end{gathered}
\end{equation}
including a Yukawa interaction term with dimensionless coupling $\lambda_t^{(5)}$. The index $``t"$ represents the fermions species considered. The most interesting case will be the one of the top quark.
We choose boundary conditions such that $t$ and $T$ have only right and left handed zero modes, respectively.    After electroweak symmetry breaking, as well as giving the zero modes mass, the Yukawa interaction induces a mixing between the different modes.  The resulting mass matrix for one flavour is of the form
\begin{equation} \label{fermionmassm}
\left(\begin{array}{cccccc}\bar{T}_L^0 & \bar{T}_L^1 &\bar{t}_L^1&\bar{T}_L^2&\bar{t}_L^2 &\hdots \end{array}\right)
\left(\begin{array}{cccccc}         m^{T,0}_{t,0} & 0 & m^{T,0}_{t,1} & 0 & m^{T,0}_{t,2} & \hdots \\ m^{T,1}_{t,0} & M_{T,1} & m^{T,1}_{t,1} & 0 & m^{T,1}_{t,2} & \hdots \\ 0 & m^{t,1}_{T,1} &M_{t,1} & m^{t,1}_{T,2} & 0 &\hdots \\  m^{T,2}_{t,0} & 0 & m^{T,2}_{t,1} & M_{T,2} & m^{T,2}_{t,2} \\ 0 & m^{t,2}_{T,1} & 0 & m^{t,2}_{T,2} & M_{t,2} & \hdots \\ \vdots & \vdots&\vdots &\vdots & \ddots          \end{array}\right)
\left(\begin{array}{c} t_R^0 \\ T_R^1 \\ t_R^1\\ T_R^2 \\ t_R^2 \\ \vdots \end{array}\right),
\end{equation}
where $M_{T,1}$ and $M_{t,1}$ are the KK masses of the doublet and singlet fields and the mixing terms are of the form
\begin{equation}
m^{\psi ,m}_{\phi ,n}=\frac{1}{\sqrt{2}}\lambda^{\psi ,m}_{\phi ,n}v_0=\frac{\lambda_t^{(5)} v_0}{\sqrt{2}L}\int_0^L dy \hspace{1mm} \sqrt{|g|} f^m_{\psi L} f^n_{\phi R} f_0.
\end{equation}
In the case of a brane Higgs, the boundary conditions imply that $m^{t,m}_{T,n}=0$ ($m,n>0$) since odd fields are zero at the IR brane.\footnote{In ref.~\cite{Azatov:2009na} the presence of such a term was argued for even in the case of a brane Higgs, once the IR brane was smeared out by regularising the delta function defining the brane and then performing an appropriate brane limit.}
With a bulk Higgs however these terms are non-zero and additional corrections arise upon diagonalization.  The mass entries $m^{t,m}_{T,n}$ vary significantly in magnitude depending on whether or not zero modes are involved, i.e.~whether $m,n=0$, and on the localisations of these zero modes. The smallest entry is $m^{t,0}_{T,0}$, which includes potential suppressions from both left and right handed zero modes. A suppression by either a left or right handed zero mode occurs for $m^{t,m}_{T,0}$ and $m^{t,0}_{T,n}$, respectively. All other entries $m^{t,m}_{T,n}$ do not suffer a suppression and therefore are of similar magnitude.

Neglecting CP violation, the mass matrix (\ref{fermionmassm}) can be partially diagonalized using orthogonal transformations of the left and right handed KK modes 
 \begin{equation}\label{fermionrot}
 O_L^TMO_R=\left(\begin{array}{ccc} 1-\frac{\theta_{L2}^2}{2} & \theta_{L1} & \theta_{L2} \\ -\theta_{L1} & 1 & 0 \\ -\theta_{L2} & 0 & 1-\frac{\theta_{L2}^2}{2}\end{array}\right)
 \left(\begin{array}{ccc} m^{T,0}_{t,0} & 0 & m^{T,0}_{t,1}  \\ m^{T,1}_{t,0} & M_{T,1} & m^{T,1}_{t,1}  \\ 0 & m^{t,1}_{T,1} &M_{t,1}  \end{array}\right)
 \left(\begin{array}{cccc} 1-\frac{\theta_{R1}^2}{2} & -\theta_{R1} & -\theta_{R2}  \\ \theta_{R1} & 1-\frac{\theta_{R1}^2}{2} & 0 \\ \theta_{R2} & 0 & 1 \end{array}\right)
 \end{equation}
where we assume a small angle approximation and consider contributions from the first KK modes only. This transformation will isolate the ``zero mode" from the KK excitations.
Below we will find that $\theta_{L1}$ and $\theta_{R2}$ are higher order in powers of $v_0/M_{KK}$, which explains the form of the orthogonal matrices used.
To find the Yukawa coupling of the physical zero mode fermion, we need to know the mixing angles in the $O_L$ and $O_R$ matrices.  Expanding to second order in powers of $v_0/M_{KK}$, we find
\begin{equation} \label{mixangs}
\begin{gathered}
\theta_{L1}\simeq-\frac{m^{T,0}_{t,0}m^{T,1}_{t,0}}{M_{T,1}^2} + \frac{m^{T,0}_{t,1}m^{t,1}_{T,1}}{M_{T,1}M_{t,1}} \hspace{2mm} ; \hspace{2mm} \theta_{L2}\simeq-\frac{m^{T,0}_{t,1}}{M_{t,1}} \\
\theta_{R2}\simeq-\frac{m^{T,0}_{t,0}m^{T,0}_{t,1}}{M_{t,1}^2}+\frac{m^{T,1}_{t,0}m^{t,1}_{T,1}}{M_{T,1}M_{t,1}} \hspace{2mm} ; \hspace{2mm} \theta_{R1}\simeq-\frac{m^{T,1}_{t,0}}{M_{T,1}}.
\end{gathered}
\end{equation}
We can see that the second terms in $\theta_{L1}$ and $\theta_{R2}$ vanish in the brane Higgs limit.
For the mass of the lowest lying mode (``zero mode") we then find 
\begin{equation}
m^{(4)}_t=m^{T,0}_{t,0}\left(1-\frac{(m^{T,0}_{t,1})^2}{2M_{t,1}^2}-\frac{(m^{T,1}_{t,0})^2}{2M_{T,1}^2}+\left(\frac{m^{t,1}_{T,1}}{m^{T,0}_{t,0}}\right)\frac{m^{T,0}_{t,1}m^{T,1}_{t,0}}{M_{T,1}M_{t,1}}+\mathcal{O}\left(\frac{m^3}{M_{KK}^3}\right)\right).
\end{equation} 
A matrix analogous to eq.~(\ref{fermionmassm}) encodes the Yukawa interactions of the fermion KK modes with the Higgs. In this matrix, diagonal terms corresponding to $M_{T,n}$ and $M_{t,n}$ are missing. This  results in a relative shift between the Yukawa coupling and mass of the ``zero mode" compared to the standard model. With the transformation defined in eq.~(\ref{fermionrot}), we find that the "zero mode" Yukawa coupling can be written as
\begin{equation}
\lambda^{(4)}_t=\lambda^{T,0}_{t,0}\left(1-\frac{3}{2}\frac{(\lambda^{T,1}_{t,0}v_0)^2}{M_{T,1}^2}-\frac{3}{2}\frac{(\lambda^{T,0}_{t,1}v_0)^2}{M_{t,1}^2}+3\left(\frac{\lambda^{t,1}_{T,1}}{\lambda^{T,0}_{t,0}}\right)\frac{\lambda^{T,0}_{t,1}\lambda^{T,1}_{t,0}v_0^2}{M_{T,1}M_{t,1}}+\mathcal{O}\left(\frac{\lambda^3v_0^3}{M^3_{KK}}\right)\right).
\end{equation}
We can now quantify the misalignment in the fermion ``zero mode" mass and Yukawa coupling as
\begin{equation} \label{rt}
r_t^{(4)}=\frac{\sqrt{2}\hspace{0.5mm}{m}^{(4)}_{t}}{{\lambda}^{(4)}_{t}v}-1=\frac{(\lambda^{T,1}_{t,0}v_0)^2}{M_{T,1}^2}+\frac{(\lambda^{T,0}_{t,1}v_0)^2}{M_{t,1}^2}-2\left(\frac{\lambda^{t,1}_{T,1}}{\lambda^{T,0}_{t,0}}\right)\frac{\lambda^{T,0}_{t,1}\lambda^{T,1}_{t,0}v_0^2}{M_{T,1}M_{t,1}}+ \frac{\delta w}{2} +\mathcal{O}\left(\frac{\lambda^3v_0^3}{M^3 _{KK}}\right).
\end{equation}
Note that because $\lambda^{T,0}_{t,1}$ is negative there is no cancellation in the contributions to $r_t^{(4)}$.
The $\delta w$ term is related to the gauge boson mass correction from section 4, eqs.~(\ref{EWlag}), (\ref{WZmass}). We use it here because the measured v.e.v., $v$, includes the mass correction to the $W$ boson.  We only include this factor for completeness since from the electroweak precision tests we know that it does only result in a negative per-mille  correction. 

Before we look at numerical evaluations of $r_t^{(4)}$, a few general statements can be made. Irrespectively of the fermion locations, $r_t^{(4)}$ scales with the 5D Yukawa couplings as $(\lambda_t^{(5)})^2$ and with the KK scale as $1/M_{KK}^2$. The third term in eq.~(\ref{rt}) is never weak in comparison to the first two terms, except for the case of a brane Higgs, where we take this term to be absent. Further statements on $r_t^{(4)}$ depend on the fermion locations. In the left-right symmetric case $c_L=-c_R$, the first and second terms in $r_t^{(4)}$ scale as $m^{(4)}_tv/M_{KK}^2$, while the third term scales as $v^2/M_{KK}^2$. So for small fermion masses the third term completely dominates. For other fermion locations these simple relations will be modified.

Our numerical evaluations of   $r_t^{(4)}$ are summarised in table \ref{results}. For the case of a bulk Higgs we use a KK scale of $M_{KK}=5.9$ TeV. 
As discussed in the previous section, a small contribution from higher dimensional operators is required in this case to reduce electroweak constraints to meet experimental bounds. For a KK scale of 8 TeV, the Yukawa deviations from the table will be reduced by a factor of $(5.9/8)^2=0.54$, while for a KK scale of 5 TeV they will increase by a factor of 1.4.
We give separate results for the three individual contributions and the total result from eq.~(\ref{rt}), $r_t^{(4)}$, denoted by (a), (b), (c) and Total, respectively. As anticipated, the third term (c) is always very important, and completely dominates for smaller fermion masses. Note that the scaling in 5D Yukawa couplings is somewhat distorted by changes in the fermion locations needed to keep the fermion mass constant. For the top quark these modifications can easily be larger than the anticipated 4\% accuracy from HL-LHC \cite{CMS:2013xfa,ATLAS:2013hta}. Also for the bottom quark the correction in the Yukawa coupling could be larger than the 2.4\% or 0.4\% accuracies aimed for at ILC and TLEP, respectively \cite{Gomez-Ceballos:2013zzn}. For the tau Yukawa coupling it seems questionable whether a deviation could be seen at ILC (predicted accuracy 2.9\%), while a detection at TLEP (predicted accuracy 0.5\%) seems promising \cite{Gomez-Ceballos:2013zzn}. 

\begin{table}[t]
\centering
\begin{tabular}{cccccccccc}
\hline
$m_t^{(4)}$ & $\lambda_t^{(5)}$ & $c_L$ & $c_R$ & $M_{T1}$ & $M_{t1}$ & (a) & (b) & (c) & Total 
\\ 
$[\rm GeV]$ & & & & [TeV] &[TeV] & [\%] & [\%] & [\%] & [\%] \\
\hline
&&&&&&&&& \\
173.48 & 4 & 0.550 & -0.26 & 6.52 & 7.12 & 12.97 & 0.05 & 19.72 & 32.7 \\
173.73 & 2 & 0.530 & -0.07 & 6.05 & 7.64 & 5.93 & 0.01 & 3.35 & 9.29 \\
173.07 & 1 & 0.488 & -0.20 & 5.98 & 7.12 & 1.29 & 0.03 & 1.31 & 2.62 \\ 
&&&&&&&&&\\ 
4.17 & 4 & 0.526 & -0.6320 & 6.04 & 6.46 & $\sim10^{-3}$ & 0.02 & 6.76 & 6.78 \\
4.17 & 2 & 0.510 & -0.6190 & 5.97 & 6.41 & $\sim10^{-3}$ & 0.02 & 2.48 & 2.50 \\
4.17 & 1 & 0.500 & -0.6004 & 5.93 & 6.33 & $\sim10^{-3}$ & 0.02 & 0.98 & 1.00 \\
&&&&&&&&&\\
1.79 & 4 & 0.542 & -0.650 & 6.10 & 6.53 & $\sim10^{-3}$ & $\sim10^{-3}$ & 3.86 & 3.87 \\
1.79 & 2 & 0.508 & 0.650 & 5.97 & 6.53 & $\sim10^{-4}$ & $\sim10^{-3}$ & 1.07 & 1.08 \\
1.79 & 1 & 0.516 & -0.621 & 6.00 & 6.41 & $\sim10^{-4}$ & $\sim10^{-3}$ & 0.58 & 0.58 \\
\end{tabular}
\caption{Relative shifts in the 4D Yukawa coupling, $r_t^{(4)}$, from eq.~(67). The columns denoted by (a), (b), (c) and Total give the first, second, third contribution and the total result in percent.
$M_{KK}$ is taken to be 5.9 TeV.}
\label{results}
\end{table}

The comparison to the case of a brane Higgs is not unique, as one has to decide which parameters should be kept constant in this procedure. In our opinion the most meaningful comparison is done by keeping the crucial off-diagonal elements $m^{T,0}_{t,1}$ and $m^{T,1}_{t,0}$ constant, in addition the to resulting 4D fermion mass. This can always be achieved by choosing a suitable brane Yukawa coupling and values for the fermion location parameters $c_L$ and $c_R$. The resulting value for $r_t^{(4)}$ can be derived from table \ref{results} by setting the contribution from column (c) to zero. The contributions from (a) and (b) will receive small changes due to the modified fermion locations. A large effect will be that for a brane Higgs we have to use a larger value of $M_{KK}=$15 TeV. So the brane Higgs cases related to the parameter sets in table \ref{results} will have values for $r_t^{(4)}$ roughly given by the sum of contributions (a) and (b) divided by a factor of four. E.g.~the brane Higgs case related to  the top quark with bulk Higgs of the first row ($r_t^{(4)}=32.7\%$) will have $r_t^{(4)}\approx(12.97\%+0.05\%)/4\approx3.3\%$. So only if the 5D Yukawa couplings is somewhat large a detection at HL-LHC seems plausible.  For lighter fermions these modifications of the Yukawa couplings will be completely undetectable in the foreseeable future.

We have numerically verified that the expressions (\ref{mixangs}) to (\ref{rt}), which are derived from considering a single KK level, receive only small corrections of $\lesssim 10\%$ when we include more fermion KK modes.

Finally we would like to remark that in variants of the warped geometry, where the KK scale is lowered to a few TeV \cite{Carena:2002dz,Agashe:2003zs,Carena:2007ua,Casagrande:2010si,Cabrer:2010si,Archer:2010bm,Archer:2010hh,Cabrer:2011fb,Cabrer:2011mw,Carmona:2011ib}, the modifications of Yukawa couplings presented in table \ref{results} have to be upscaled accordingly. In the case of a KK scalar of 3 TeV, the corrections from in table \ref{results} will increase by a factor of 3.9. Then 5D top Yukawa couplings $\lambda_t^{(5)}\gtrsim1.5$ will then already be disfavoured by present observations of Higgs production via gluon fusion at the LHC. 

Yukawa coupling misalignment also has impact on flavour violation mediated by Higgs exchange, as e.g.~discussed in \cite{Casagrande:2008hr,Buras:2009ka,Azatov:2009na}. Also Higgs corrections to the muon anomalous magnetic moment were found to depend on the Higgs localisation \cite{Beneke:2014sta}.
Analysing the resulting constraints for the scenario investigated here, however, is beyond the scope of the present work. For generic anarchic Yukawa couplings Higgs induced flavour violation will be large, certainly pushing the bound on the KK scale beyond the bounds we derived from electroweak precision constraints in section 4. However, flavour violation can be significantly reduced if fermion localizations are generation independent (at least for the first and second generation). In such a case the dominant bounds on the KK scale are those we derived in section 4. Here we conclude that unavoidable Yukawa misalignment does not lead to additional bounds on the KK scale.

\section{Conclusions}

In this paper we have revisited the scenario of a bulk Higgs in warped extra dimensions, without assuming deviations from AdS space or imposing a custodial symmetry. Our aim is to investigate the robustness of bounds on the KK scale from electroweak observables and modifications of SM Yukawa couplings. We then discuss prospects for observing new physics at future collider experiments. 

Performing a standard electroweak precision analysis, we confirm that a bulk Higgs brings down the limit on the KK scale, which we take to be mass of the lightest vector resonance, form about 15 TeV to about 8 TeV. A bulk Higgs reduces impact of KK gauge boson excitations on the SM particles after electroweak symmetry breaking. The Higgs, being a bulk field, also has KK excitiations which contribute to gauge boson masses etc., but their impact is unobservable for the foreseeable future. However, deviations from the SM values of the HZZ and HWW couplings at the sub-percent level will be induced by KK gauge boson mixing. These effects will be very difficult to see at ILC, but TLEP with a predicted sensitivity of better than 0.2\% could detect them. 

We then include into the analysis higher dimensional operators which parametrize unknown contributions from the UV completion of our setup. We find that a dimension-8 operator in 5D can have an non-negligible impact on the T-parameter. The bound on the KK scale of 8 TeV, derived previously is therefore not robust. We therefore argue that this unknown contribution could bring the KK scale down to at least about 5 TeV. The LHC run at 13 TeV could then discover KK resonances in the simple scenario presented here.

Finally, we investigate whether additional bounds on the KK scale can be derived from deviations in fermion Yukawa couplings, in particular for the top quark. We find that even with a KK scale of only 5 TeV, the enhancements in the top Yukawa coupling can be larger than $10\%$.  However there are areas of parameter space where this enhancement can be much smaller and hence this will not generally lead to tension with observed Higgs production at LHC8. Such a tension would require large values of the associated 5D Yukawa coupling. In the future it will be interesting to look for Yukawa deviations for that quark at high-luminosity LHC. The enhancements in the bottom and tau Yukawa couplings can also be as large as a few percent, making this detectable at ILC and TLEP.
Furthermore, top Yukawa coupling misalignment should be taken into account in models where top loops induce electroweak symmetry breaking, e.g. warped geometry realisations of composite Higgs models \cite{Agashe:2004rs,Contino:2006nn,Contino:2010rs}.

As is well known, models of the type presented here often generate large flavour and CP-violation from KK exchange. These may induce bounds on the KK scale, which are much more stringent than the ones we have considered. However, one should bare in mind that these flavour bounds depend on how the fermion mass pattern is generated, and can be reduced or almost avoided by flavour symmetries.

So we conclude that, even without an enlarged gauge symmetry, a bulk Higgs in pure AdS space opens the possibility to discover KK resonances at the upcoming LHC run.
\newpage

\section*{Acknowledgements}
We would like to thank Veronica Sanz for useful discussions and comments.  B.M.D. was supported by the Science and Technology Facilities Council.  S.J.H. is supported by the Science Technology and Facilities Council (STFC) under grant ST/J000477/1.

\bibliography{BibTeXReferenceList2}{}
\bibliographystyle{h-physrev3}

\end{document}